\documentclass[twocolumn,prb,showpacs]{revtex4}
\usepackage{graphicx}
\usepackage{amssymb}

\newcommand{\GSO}{$\rm Gd_2Sn_2O_7$}
\newcommand{\GTO}{$\rm Gd_2Ti_2O_7$}

\begin{document}

\title{Magnetic excitations in dipolar pyrochlore
antiferromagnet Gd$_2$Sn$_2$O$_7$}

\author{S. S. Sosin}
\affiliation{P. L. Kapitza Institute for Physical Problems RAS,
119334 Moscow, Russia \\
Commissariat \`{a} l'Energie Atomique,
DSM/INAC/SPSMS, 38054 Grenoble, France}

\author{L. A. Prozorova}
\affiliation{P. L. Kapitza Institute for Physical Problems RAS,
119334 Moscow, Russia}

\author{P. Bonville}
\affiliation{Commissariat \`{a} l'Energie Atomique,
DSM/IRAMIS/SPEC, 91191 Gif sur Yvette, France}

\author{M. E. Zhitomirsky}
\affiliation{Commissariat \`{a} l'Energie Atomique,
DSM/INAC/SPSMS, 38054 Grenoble, France}

\date{\today}

\begin{abstract}
The spin dynamics in the geometrically frustrated pyrochlore
antiferromagnet $\rm Gd_2Sn_2O_7$ is studied by means of the electron
spin resonance. In the ordered phase ($T_N = 1$\,K), we have
detected three gapped resonance modes. Their values agree well
with the developed spin-wave theory which takes into account the
Heisenberg nearest-neighbor exchange, the single-ion anisotropy
and the long-range dipolar interactions. The theory also predicts
a fourth lowest-frequency gap, which lies beyond the experimental
range of frequencies, but determines the exponential decrease of
the specific heat at low temperature.
\end{abstract}


\maketitle

\section{Introduction}

The geometrically frustrated  Heisenberg antiferromagnet on a
pyrochlore lattice has an extensive degeneracy of the ground
state.  \cite{moessner98} As a result, weak additional
interactions play a prominent role leading to a multitude of
magnetic phases and phenomena. A typical example is provided by
\GSO\ and \GTO, two pyrochlore antiferromagnets with close values
of the nearest-neighbor exchange constants, which, nevertheless,
have different ordered magnetic structures. While  the
antiferromagnetic structure in \GTO\ is described by the
ordering wave-vector ${\bf q} = (1/2,1/2,1/2)$ and has not yet
been completely resolved, \cite{stewart} the second material \GSO\
has a much simpler four-sublattice spin-structure with ${\bf q} = 0$. \cite{wills}

An exponential decrease of the low temperature specific heat of
\GSO\ was recently found by the calorimetric  measurements.
\cite{gingras2} This result unambiguously points at a gapped
excitation spectrum in the ordered magnetic phase. It is somewhat
unexpected in view of the results of analogous measurements for
\GTO, \cite{yaouanc} which yield a power-law behavior $C\propto
T^2$ down to temperatures of the order of 100~mK. Presence of
non-frozen magnetic degrees of freedom at very low
temperature was also related to the persistent spin dynamics
observed via the muon spin relaxation ($\mu$SR) in \GTO
\cite{yaouanc,dunsiger} and via M\"ossbauer \cite{bertin} and
$\mu$SR measurements \cite{dalmas,bonv1} in \GSO. The observed
difference in the low-temperature dependence of the specific heat in
the two pyrochlore antiferromagnets must be further
corroborated by a direct investigation of the
excitation spectrum. Since inelastic neutron scattering
measurements are hindered in Gd-compounds, the most convenient
experimental technique left is the electron spin resonance (ESR).
An absorption of photons with a typical wave-length $\lambda\sim 0.1
\div 10$~cm provides a high-resolution probe of
low-energy excitations. For the most probable single photon-magnon
process only magnons with small wave-vectors $k\sim
1/\lambda\rightarrow 0$ can be excited.

Our previous measurements of the ESR spectra in \GSO\ were performed
at temperatures between 10~K (the Curie-Weiss constant is $\vert
\theta_{CW} \vert \simeq 9$~K) and the ordering transition at $T_N
= 1$~K.\cite{sosin3} This temperature range is characterized by
strong spin correlations in the absence of an order parameter,
which is commonly called a cooperative paramagnet. An unusual
transformation of an exchanged narrowed paramagnetic line into a
single gapped resonance mode with linear field dependence was
observed in this regime.

The present work investigates further transformations of the
resonance spectrum at temperatures below the magnetic ordering
transition. Three gapped modes are observed, two of which are
degenerate in zero magnetic field. The gap values are reproduced
by the spin-wave calculations using the known values of the exchange constant,
the dipolar and the single-ion anisotropy energies. The fourth
gap predicted by the theory remains unobserved in our experimental
range, but agrees well with the specific heat data.

\section{Samples}

Powder samples of \GSO\ were prepared by the method described in a
previous publication. \cite{bonville} For sample characterization
we have measured the specific heat in a $^3$He-$^4$He dilution
fridge in the temperature range 0.1--2~K. The data are obtained
using a quasi-adiabatic technique with continuous heating of the
sample. \cite{calem} In order to improve the thermal contact
between the insulating powder sample and the sample holder, 1.5~mg
of the powder was mixed with 1~mg of Apiezon~N grease, wrapped
into a silver foil and pressed. The quality of the thermal contact
has been verified by comparing the data obtained in experiments
with different heating rates. The range of fully reliable results
corresponds to temperatures above 0.15~K.

\begin{figure}
\centerline{\includegraphics[width=\columnwidth]{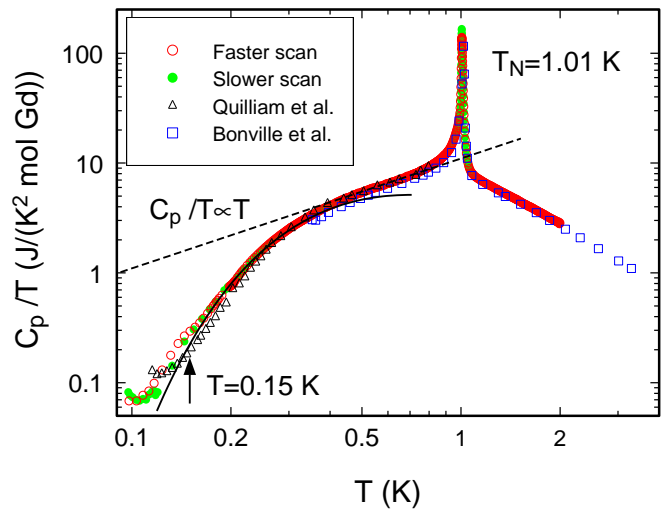}}
\caption{(Color online.) Thermal variation of the specific heat
divided by temperature in \GSO\ measured with different heating
rates ({\Large $\bullet$} and {\Large $\circ$} are slower and
faster scans respectively); $\square$ and $\vartriangle$ are
previous data from Refs.~\protect\onlinecite{bonville,gingras2};
the dashed line is a linear approximation to the high temperature
part of the data, the fit by a solid line is described in the
text, the arrow marks the low temperature limit of data
reliability.} \label{fig1}
\end{figure}

Our experimental results for the specific heat are presented in
Fig.~\ref{fig1}. Two subsequent temperature scans shown by open
and closed circles (slower and faster scans respectively) are in
satisfactory agreement with each other. A sharp peak is observed
at the ordering transition with $T_N=1.01$~K. The previously
published data \cite{bonville} obtained with the same measurement
technique are shown by squares and lie 5--10\% lower in the whole
range of temperatures. The measurements from
Ref.~\onlinecite{gingras2} (triangles) demonstrate a perfect
agreement with our data above 0.4~K, where both curves can be
empirically fitted by a $C\propto T^2$ dependence. Both sets of
data also agree, but less perfectly, below 0.4~K and show a
drastic decrease on cooling. According to our measurements, the
most significant reduction of the specific heat (by a factor of
20) occurs in the temperature interval 0.3--0.15~K.
Decreasing temperature further down to 0.1~K leads to
an additional specific heat reduction by a factor of 5 to the level,
which does not exceed the experimental accuracy.
The specific heat decreases faster than any
reasonable power law $T^n$, which suggests an exponential
temperature dependence in agreement with the previous study.
\cite{gingras2}

\section{Magnetic resonance}

\begin{figure}
\centerline{\includegraphics[width=\columnwidth]{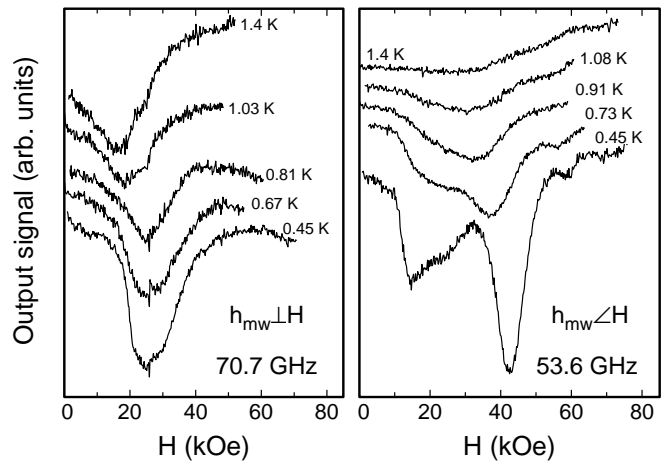}}
\caption{Evolution of the magnetic resonance absorption spectra in
a powder sample of \GSO\ on cooling from above the ordering
transition to $T=0.45$~K, recorded in a perpendicular polarization
of the microwave field with respect to the external field ${\bf
h}_{\rm mw}\perp {\bf H}$ (left panel) and in a tilted
polarization (right panel); the recorded lines are shifted upwards
for clarity.} \label{fig2}
\end{figure}

Magnetic resonance measurements have been carried out in a
transmission type spectrometer with a cylinder cavity designed for
frequencies above 25~GHz, which was equipped with a $^3$He
cryostat with a minimum working temperature of 0.4~K. The magnetic
field up to 100~kOe is generated by a cryomagnet. The absorption
spectra are recorded on forward and backward field sweeps.

To start with, we have traced the evolution of the resonance
absorption lines on cooling the sample from the strongly
correlated disordered state through the ordering transition at
$T_N=1.0$~K down to the lowest experimental temperature (0.45\,K),
at which the system is fully ordered (the magnetic transition is
first order \cite{bonville}). Two sets of measurements with
different polarizations of the microwave field with respect to the
external magnetic field were performed. The left panel of
Fig.~\ref{fig2} shows the resonance spectra of a sample glued onto
the bottom of the cavity where the microwave field has only a
component perpendicular to the external field ${\bf h}_{\rm
mw}\perp {\bf H}$. The single resonance line observed at all
frequencies for $T>T_N$ changes its shape and shifts to larger
fields when going through the transition. This shift results from
an unusual linear field dependence of the resonance gap in the
cooperative paramagnetic state, as discussed in
Ref.~\onlinecite{sosin3}.

The absorption spectra are significantly modified when the sample
is placed into a microwave field with a component along the
external field. Additional lines develop in the spectrum below
1~K, one of them having a much larger intensity than the others.
The properties of all these resonance modes were studied in detail
at the lowest experimental temperature 0.45~K.

\begin{figure}
\centerline{\includegraphics[width=\columnwidth]{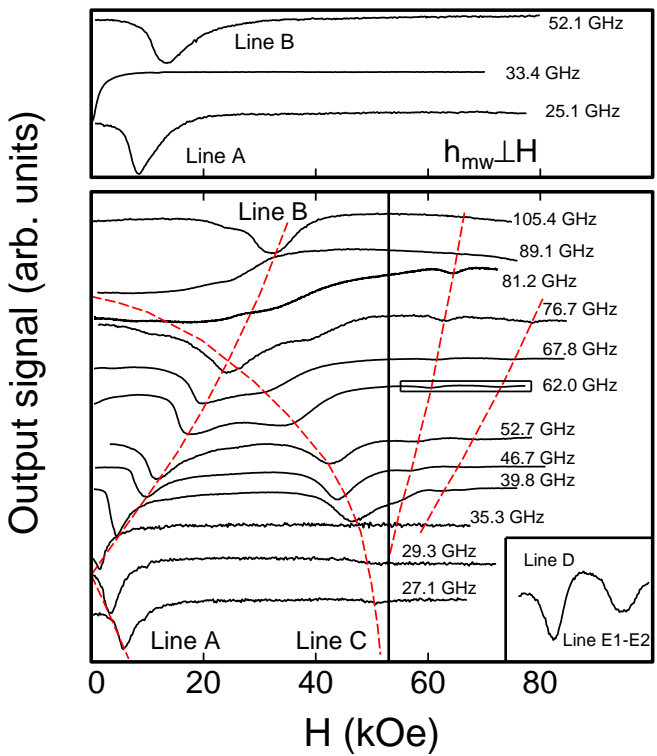}}
\caption{(Color online.) The magnetic resonance absorption spectra
in a powder sample of \GSO\ recorded at $T=0.45$~K for various
frequencies with ${\bf h}_{\rm mw}\perp {\bf H}$ (upper panel) and
in a tilted direction of ${\bf h}_{\rm mw}$ (lower panel); dashed
lines are guide-to-eye for tracing the field evolution of
different resonance lines labeled by letters A-E; the inset in the
lower panel expands the in-frame part of the absorption record at
$\nu =62.0$~GHz including spectral lines D and E. The transition
into a spin-polarized phase at $H_s=53$~kOe is marked by a
vertical line.} \label{fig3}
\end{figure}

The upper panel of Fig.~\ref{fig3} shows the resonance spectra for
a sample with ${\bf h}_{\rm mw}\perp {\bf H}$. These data have
been briefly discussed before. \cite{sosin2} The single resonance
lines observed at various frequencies belong to two different
branches, one of which is decreasing (line~A) and the other one is
increasing (line~B) in field. The extrapolation of these lines to
zero field gives the same gap value of $33.5\pm 0.5$~GHz ($1.61\pm
0.02$~K) for both branches, which points to an exact degeneracy of
the corresponding magnetic excitations. No other resonance
absorption was detected within the experimental accuracy of 0.5\%.

The absorption spectra obtained for a tilted direction of ${\bf
h}_{\rm mw}$ are presented on the lower panel of Fig.~\ref{fig3}.
The intense sharp resonance line shown in Fig.~\ref{fig2}, which
arises in the ordered phase, appears to be a third branch with a
gap of $85\pm 5$~GHz ($4.1\pm 0.2$~K, line~C). This mode decreases
with increasing magnetic field and softens in the vicinity of the
transition into the fully saturated phase. Extrapolating its field
dependence to zero frequency, one can determine the value of the
saturation field $H_s = 53.0 \pm 0.5$~kOe. A careful study of the
spectrum above the critical field again reveals the existence of
three weak components (line~D and double split line~E) increasing
in field, with the line~E doublet exhibiting an exact linear field
dependence.

Our ESR measurements directly detect three resonance modes in zero
magnetic field. Two of them (lines~A and B) are exactly degenerate
with a gap $\Delta_{2,3}=1.61$~K and the third mode has a larger
gap of $\Delta_4=4.1$~K. In addition, the previous specific heat
measurements and theoretical calculations suggest the presence of
a fourth lowest branch at energy $\Delta_1 \sim  1.2$~K.
\cite{gingras2,gingras1} The corresponding frequency $\nu_1 \sim
25$~GHz lies near the lower boundary of our experimentally
accessible range. Hence, it is  natural that this mode remains
unobserved in the ESR experiment provided that it decreases with
increasing magnetic field. A crude estimate from the low-$T$ fit
of our specific heat data $C(T) \propto T^{-1/2} e^{-\Delta/T}$
(the corresponding fit is shown by the solid line on
Fig.~\ref{fig1}) also gives $\Delta_1 \sim 1.0$~K. One should note
here that more elaborate fits at temperatures below 0.15~K
(including, {\it e.g.}, nuclear contributions) would rely on an
overestimation of the experimental accuracy due to the degradation
of the thermal contact in insulating powder samples in this
temperature range.

The previous spin-wave calculations \cite{gingras1} reflect the
general features of the measured spectrum, with a 10--20\%
accuracy for the observed gaps. The principal qualitative
difference with our work is that our ESR measurements find two
{\it exactly} degenerate magnon branches at ${\bf k}=0$ in zero
magnetic field, whereas Ref.~\onlinecite{gingras1} predicts a
finite splitting between them: $\Delta_2=1.76$~K and
$\Delta_3=1.93$~K. In the next section we present the detailed
theoretical calculations of the ESR spectra in \GSO, which not
only yield the correct degeneracy of ${\bf k} = 0$ magnons but
also show an overall improved agreement with the experimental data.

\section{Spin-wave theory}

The unit cell of the pyrochlore lattice contains four magnetic
atoms. Their positions and the specific choice of the local axes
adopted below is
\begin{eqnarray}
&& \mbox{\boldmath $\rho$}_1 = (0,0,0)\ , \quad \ \ \; \hat{\bf
z}_1 = \frac{1}{\sqrt{2}}(1,-1,0)\ , \nonumber \\ &&
\mbox{\boldmath $\rho$}_2=\Bigl(0,\frac{1}{4},\frac{1}{4}\Bigr)\ ,
\quad \hat{\bf z}_2 = \frac{1}{\sqrt{2}}(-1,-1,0)\ , \nonumber \\
&& \mbox{\boldmath
$\rho$}_3=\Bigl(\frac{1}{4},0,\frac{1}{4}\Bigr)\ , \quad \hat{\bf
z}_3 = \frac{1}{\sqrt{2}}(1,1,0)\ , \nonumber \\ &&
\mbox{\boldmath $\rho$}_4 =  \Bigl(
\frac{1}{4},\frac{1}{4},0\Bigr)\ , \quad \hat{\bf z}_4 =
\frac{1}{\sqrt{2}}(-1,1,0) \label{axes}
\end{eqnarray}
with $\hat{\bf y}_i \equiv (0,0,1)$ and $\hat{\bf x}_i = \hat{\bf
y}_i \times\hat{\bf z}_i$. The unit cell tetrahedra form a
face-centered cubic structure with the translational basis ${\bf
a}_1 = (0,\frac{1}{2},\frac{1}{2})$, ${\bf a}_2 =
(\frac{1}{2},0,\frac{1}{2})$, ${\bf a}_3 =
(\frac{1}{2},\frac{1}{2},0)$.

In the following we will take into account all major magnetic
interactions starting with the strongest nearest-neighbor
exchange:
\begin{equation}
\hat{\cal H} = J \sum_{\langle ni,mj\rangle}{\bf S}_{ni} \cdot
{\bf S}_{mj} \ , \label{H1}
\end{equation}
where $n,m$ denote unit cells and $i,j=1$--4 indicate position
inside cell. The equilibrium magnetic structure determined for
\GSO\ in neutron diffraction experiments \cite{wills} is a
four-sublattice chiral spin-cross with ${\bf S}_i\parallel
\hat{\bf z}_i$, which is also known as the Palmer-Chalker state.
\cite{palmer} This state is one out of the many degenerate spin
structures obeying the classical constraint: $\sum_i {\bf S}_i =
0$ for every tetrahedron, which must be fulfilled by the ground
state in the absence of single-ion anisotropy and dipolar
interactions.

Let us first calculate the excitation spectrum for the chiral
spin-cross state neglecting all magnetic anisotropies. The local
coordinate system is used for the spin operators ${\bf S}_i =
S_i^x \hat{\bf x}_i + S_i^y\hat{\bf y}_i + S_i^z \hat{\bf z}_i$,
where the three axes are defined above, Eq.~(\ref{axes}). Since we
are dealing with the ${\bf k}=0$ magnetic structure, the local
axes do not carry the cell index $n$. The Holstein-Primakoff
transformation is used for the bosonic representation of spin
operators. Since gadolinium spins $S=7/2$ are large, we neglect
interactions between magnons and always keep up to the quadratic
terms in the bosonic Hamiltonian. With the chosen accuracy it is
sufficient to write
\begin{eqnarray}
&& S_i^z = S - a^\dagger_i a_i \ , \ \ \ S_i^x =
\sqrt{\frac{S}{2}}\,(a^\dagger_i + a_i) \ , \nonumber \\ && S_i^y
= \sqrt{\frac{S}{2}}\, i(a^\dagger_i - a_i) \ . \label{HP2}
\end{eqnarray}
The obtained quadratic form is diagonalized by performing
consecutively Fourier and Bogoliubov transformations. The
excitation spectrum consists of four branches:
\begin{equation}
\varepsilon_{1,2}({\bf k}) \equiv 0\ , \ \ \ \varepsilon_{3,4}
({\bf k}) = 2JS \sqrt{1 - \cos\frac{k_x}{2} \cos\frac{k_y}{2} } \
.
\end{equation}
Two of them have zero energies everywhere in the Brillouin zone,
which reflects the infinite degeneracy of the nearest-neighbor
exchange Hamiltonian on a pyrochlore lattice. Interestingly, the
two other branches have purely two-dimensional dispersions in the
harmonic approximation. Various perturbations to the
nearest-neighbor Heisenberg Hamiltonian (\ref{H1}) should produce
a finite dispersion of the two lowest branches, but the quasi 2D
behavior of $\varepsilon_{3,4} ({\bf k})$ for the two
higher-energy modes should survive in a certain range of
parameters.

As the next step we add the single-ion anisotropy to (\ref{H1}) in
the form of the lowest-order crystal-field harmonics:
\begin{equation}
\hat{\cal H}_a = D_a \sum_i \bigl({\bf S}_{i} \cdot {\bf n}_i
\bigr)^2 \ ,
\end{equation}
${\bf n}_i$ being the four local anisotropy axes parallel to the
principal cubic diagonals. Keeping again only quadratic terms in
the bosonic representation one obtains:
\begin{eqnarray}
({\bf S}_i \cdot {\bf n}_i)^2 & = & S \Bigl( a_i^\dagger a_i +
\frac{1}{2}\Bigr)  \\ && \mbox{} + \frac{S}{6} \Bigl[ a_i^2
(1\pm2\sqrt{2}i) + a_i^{\dagger 2}(1\pm2\sqrt{2}i) \Bigr] \ ,
\nonumber
\end{eqnarray}
where the upper (lower) sign corresponds to $i = 1,4$ (2,3).

Since we are interested in the ESR spectrum given by ${\bf k}=0$
magnons, we simplify the following calculations by considering
only uniform modes, avoiding hence a step with the Fourier
transformation. The exchange Hamiltonian projected onto the
four-site magnetic unit cell is written as
\begin{eqnarray}
&& \hat{\cal H}^{(2)}_1/JSN = 2 (a_1^\dagger a_1 + a_2^\dagger a_2
+ a_3^\dagger a_3 + a_4^\dagger a_4)  \nonumber \\ & &
\mbox{}\qquad\quad +  a_1^\dagger(a_2+a_3) + a_2^\dagger(a_1+a_4)
 +  a_3^\dagger(a_1+a_4)
\nonumber \\ & & \mbox{}\qquad\quad
 + a_4^\dagger(a_2+a_3)
 - a_1(a_2 + a_3 + 2a_4)
\nonumber \\ & & \mbox{}\qquad\quad - a_2(2a_3 + a_4) - a_3 a_4 +
{\rm h.\,c.}\,,
\end{eqnarray}
where $N$ is the number of unit cells in the lattice. The
single-ion anisotropy is represented as
\begin{eqnarray}
&& \hat{\cal H}_a^{(2)}/D_aSN = a_1^\dagger a_1 + a_2^\dagger a_2
+ a_3^\dagger a_3 + a_4^\dagger a_4
\\ & & \mbox{} \qquad\quad +
\frac{1}{6}(1-2\sqrt{2}i) (a_1^2 + a_4^2 + a_2^{\dagger 2} +
a_3^{\dagger 2}) + {\rm h.\,c.} \nonumber
\end{eqnarray}
Diagonalization of quadratic forms with the help of a generalized
Bogoliubov transformation has been described many times in the
literature and will not be repeated here. In the present case the
diagonalization can be performed analytically yielding four magnon
gap energies:
\begin{equation}
\Delta_1 = 0\ , \quad \Delta_{2,3} = 2S\sqrt{2D_aJ/3}\ , \quad
\Delta_4 =  2\Delta_2 \ .
\end{equation}
The lowest magnon branch remains gapless in spite of the
single-ion term. This is in agreement with the analysis of
Ref.~\onlinecite{champion04}, which finds an infinite (but not
extensive) degeneracy for the easy-plane pyrochlore
antiferromagnet.

The double degeneracy of the two intermediate magnon modes
$\Delta_{2,3}$ follows from the tetragonal symmetry of the chiral
spin-cross structure. Analysis of the eigenvectors of the
Bogoliubov transformation identifies corresponding oscillations
with {\it out-of-plane} motion of only one pair of spins $S_1,S_4$
or $S_2,S_3$ with opposite phases. The two modes transform into
each other according to the 2D irreducible representation of the
tetragonal point group. Extra interactions (dipolar, etc.) will
not modify such a degeneracy as long as the chiral spin-cross
structure remains stable. The two other  modes $\Delta_{1,4}$
correspond to predominantly {\it in-plane} motion of all four
spins.

For a simple numerical estimate we use the following parameters
for ${\rm Gd_2Sn_2O_7}$: $\vert \theta_{\rm CW} \vert= 2JS(S+1)
\approx 8.6$~K \cite{bonville} with $J = 2/63\ \vert \theta_{\rm CW} \vert
= 0.27$~K and $D_a \approx 0.14$~K. \cite{glazkov} This
yields the following values $\Delta_{2,3} = 1.12$~K and $\Delta_4
= 2.24$~K, which are somewhat lower than the experimentally
measured frequencies (1.61 and 4.1\,K respectively). Introduction
of further neighbor exchanges will not modify the above results:
(i) the third-neighbor exchange $J_3$ (the notation is taken from
Ref.~\onlinecite{wills}) couples spins on the same sublattice and,
consequently, does not contribute to the ${\bf k}=0$ modes, (ii)
the antiferromagnetic second-neighbor exchange $J_2$ yields the
same replacement $J \rightarrow J+2J_2$ in the expressions for
$\theta_{CW}$ and for the uniform modes and does not, therefore,
change the gaps.

The next important perturbation to the exchange energy (\ref{H1})
is the dipole-dipole interaction: \cite{gingras1,gingras1b}
\begin{equation}
\hat{\cal H}_{\rm dip} = \frac{D}{2S^2}\!\! \sum_{ni,mj}\!\!
\frac{{\bf S}_{ni}\cdot {\bf S}_{mj} - 3({\bf S}_{ni} \cdot {\bf
r}^{ij}_{nm}) ({\bf S}_{mj} \cdot {\bf r}^{ij}_{nm}) }{|{\bf
R}^{ij}_{nm}|^3} \label{Hdip}
\end{equation}
with $D = (g\mu_BS)^2/a^3$, ${\bf R}^{ij}_{nm}$ being the vector
linking two spins, measured in units of the lattice constant $a$,
and ${\bf r}^{ij}_{nm} = {\bf R}^{ij}_{nm}/|{\bf R}^{ij}_{nm}|$.
The strength of the dipolar coupling between two neighboring spins
in \GSO\ is estimated as
\begin{equation}
E^d_{\rm n.n.} = \frac{(g\mu_B S)^2}{(a/2\sqrt{2})^3} =
16\sqrt{2}\,D = 0.605\ {\rm K} \ ,
\end{equation}
where we have substituted $a = 10.455$\AA\ for the lattice
constant. \cite{kennedy} The parameter $E^d_{\rm n.n.}$ is
three-times smaller than the single-ion energy $D_a S^2 \sim
1.73$~K, but the dipolar interactions still play an important role
due to their long-range nature. From now on we define
dimensionless parameters $d_a = D_a/J= 0.516$ and $d = D/JS^2 =
0.008$ normalizing all interactions to the exchange constant $J$
and the excitation energies to $JS$.

Projecting (\ref{Hdip}) onto the four-sublattice magnetic
structure (${\bf S}_{ni} \equiv {\bf S}_i$) we obtain
\begin{equation}
\hat{\cal H}_{\rm dip} = \frac{1}{2} N \sum_{ij,\alpha\beta}
S_{i}^{\alpha}S_{j}^{\beta} D _{ij}^{\alpha\beta} \ ,
\label{Hdip1}
\end{equation}
where the dipolar matrix is given by
\begin{equation}
D_{ij}^{\alpha\beta} = d \sum_m \frac{1}{|R^{ij}_{nm}|^3}
\Bigl[\delta_{\alpha\beta} - 3({\bf r}^{ij}_{nm})^\alpha ({\bf
r}^{ij}_{nm})^\beta \Bigr]
 \ .
\end{equation}
In a cubic crystal the diagonal matrix elements are
$D_{ii}^{\alpha\beta} \sim \delta_{\alpha\beta}$ and drop out from
the equations on the equilibrium spin configuration and on the
energies of the ${\bf k} = 0$ modes.

The dipolar sums are straightforwardly evaluated using the Ewald's
summation technique:\cite{ewald}
\begin{widetext}
\begin{eqnarray}
D_{ij}^{\alpha\beta }/d & = & 16\pi \sum_{\bf G}{'}\;
\frac{G^\alpha G^\beta}{G^2}\: e^{-G^2/4Q_c^2}\: e^{i{\bf G}\cdot
(\mbox{\boldmath\scriptsize $\rho$}_i - \mbox{\boldmath\scriptsize
$\rho$}_j)} + \sum_{\bf R} \biggl\{\, \textrm{erfc}(Q_cR_{ij})
\biggl[\frac{\delta_{\alpha\beta}}{R_{ij}^3}  -
\frac{3R_{ij}^\alpha R_{ij}^\beta} {R_{ij}^5}\biggr] \nonumber \\
&& \mbox{} - \frac{2Q_c}{\sqrt{\pi} R_{ij}^2}\: e^{-Q_cR_{ij}^2}\:
\biggl[2Q_c^2 R_{ij}^\alpha R_{ij}^\beta - \delta_{\alpha\beta} +
\frac{3R_{ij}^\alpha R_{ij}^\beta} {R_{ij}^2} \biggr] \biggr\} \ ,
\label{Dip_sums}
\end{eqnarray}
\end{widetext}
where $Q_c\sim 1$ is an arbitrary cutoff, ${\bf G}$ is a
reciprocal lattice vector, ${\rm erfc}(x)$ is the complementary
error function, and ${\bf R}_{ij} = {\bf R} +
\mbox{\boldmath$\rho$}_i - \mbox{\boldmath$\rho$}_j$, $\bf R$
being an fcc lattice vector. Summations in Eq.~(\ref{Dip_sums})
are performed over all $\bf R$ and $\bf G$ excluding ${\bf G}=0$.

Cubic symmetry of the pyrochlore lattice leaves only three
independent constants
\begin{eqnarray}
&& D_{12}^{xx} = D_{34}^{xx} = c_1 \ , \quad D_{12}^{yy} =
D_{12}^{zz} = D_{34}^{yy} = D_{34}^{zz} = c_2 \ , \nonumber \\ &&
D_{12}^{yz} = D_{12}^{zy} = -D_{34}^{yz} = -D_{34}^{zy} = c_3 \ ,
\nonumber \\ && D_{13}^{yy} = D_{24}^{yy} = c_1 \ , \quad
D_{13}^{xx} = D_{13}^{zz} =D_{24}^{xx} =D_{24}^{zz} = c_2 \ ,
\nonumber \\ && D_{13}^{xz} = D_{13}^{zx} = -D_{24}^{xz} =
-D_{24}^{zx} = c_3 \ , \nonumber \\ && D_{14}^{zz} = D_{23}^{zz} =
c_1 \ , \quad D_{14}^{xx} = D_{14}^{yy} = D_{23}^{xx} =
D_{23}^{yy} = c_2 \ , \nonumber \\ && D_{14}^{xy} = D_{14}^{yx} =
-D_{23}^{xy} = -D_{23}^{xy} = c_3 \ .
\end{eqnarray}
Evaluating numerically the corresponding expressions
(\ref{Dip_sums}) we find $c_1 = 17.92 d = 0.143$, $c_2 = -34.09 d
= -0.273$, and $c_3 = -57.84 d = -0.463$.

The dipolar energy (\ref{Hdip1}) can be used to compare the
relative stability of different ${\bf q} = 0$ magnetic structures.
In particular, the dipolar contribution for the chiral spin-cross
configuration (\ref{axes}), or the Palmer-Chalker state,  is
$E^d_1/NS^2 = -2c_1 + 2c_3$. The alternative (non-chiral)
spin-cross structure, which is realized in Er$_2$Ti$_2$O$_7$
\cite{champion03} and can be obtained from the chiral spin cross
with ${\bf S}_2 \rightarrow -{\bf S}_2$, ${\bf S}_3 \rightarrow
-{\bf S}_3$, has a higher dipolar energy $E^d_2/NS^2 = 2c_1 - 4c_2
+ 2c_3$.

To calculate the effect of the dipolar interaction on the magnon
spectra we transform again to the local spin frame in
Eq.~(\ref{Hdip1}) and bosonize spin operators using (\ref{HP2}).
The dipolar matrix elements in the rotating coordinate system
${\cal D}_{ij}^{\alpha\beta}$ are expressed via the laboratory
frame matrix $D_{ij}^{\alpha\beta}$ by
\begin{equation}
{\cal D}_{ij}^{\alpha\beta} = \hat{\bf e}_{i\alpha}^\mu \hat{\bf
e}_{j\beta}^\nu D_{ij}^{\mu\nu} \ , \label{transform}
\end{equation}
where $\hat{\bf e}_{i\alpha}$ are the local basis vectors, see
Eq.~(\ref{axes}).

The obtained quadratic form of bosonic operators is
\begin{widetext}
\begin{eqnarray}
\hat{\cal H}^{(2)}_{\rm dip} & = & \sum_{\langle ij \rangle} -
{\cal D}_{ij}^{zz}(a^\dagger_i a_i + a^\dagger_j a_j) +
\frac{1}{2} ({\cal D}_{ij}^{xx}+{\cal D}_{ij}^{yy}) (a^\dagger_i
a_j + a^\dagger_j a_i) - \frac{i}{2} ({\cal D}_{ij}^{xy}+{\cal
D}_{ij}^{yx}) ( a^\dagger_i a_j - a^\dagger_j a_i) \nonumber \\ &&
\mbox{} + \frac{1}{2}({\cal D}_{ij}^{xx}-{\cal D}_{ij}^{yy}) (a_i
a_j + a^\dagger_i a^\dagger_j) - \frac{i}{2}({\cal D}_{ij}^{xy}
+{\cal D}_{ij}^{yx}) ( a_i a_j - a^\dagger_i a_j^\dagger) \ .
\end{eqnarray}
\end{widetext}
The summation is performed over all sublattice pairs. Skipping the
straightforward algebra behind the substitution (\ref{transform})
and the subsequent Bogoliubov transformation, we present the final
results for the magnon energies. The two degenerate modes have the
energy:
\begin{equation}
\varepsilon^2_{2,3} = \frac{1}{3} \bigl[ (2d_a-3c_3)\,(4+c_1+c_2)
  - 4c_3d_a \bigr] \ ,
\end{equation}
while the energies of the two other modes are given by (positive)
roots of
\begin{eqnarray}
&& 3\varepsilon^4  - 2 \varepsilon^2\bigl[6(c_1-c_2)(4+c_1+c_2)
\\ & & \mbox{} \quad +(2d_a-3c_3)(8+5c_1-c_2-4c_3)\bigr]
\nonumber\\ & & \mbox{} \quad +
8(c_1-c_2)\bigl[2d_a-3c_3+2(c_1-c_2)\bigr] \nonumber\\ & & \mbox{}
\quad \times \bigl[(4+c_1+c_2)(2d_a-3c_3)-4c_3d_a \bigr] = 0 \ .
\nonumber
\end{eqnarray}
Restoring the scaling parameter $JS$ and using the following
values for the microscopic parameters: $J = 0.27$~K, $d_a =
0.516$, and  $d = 0.008$ we obtain for the gaps: $\Delta_1 =
1.24$~K, $\Delta_{2,3} = 1.77$~K, $\Delta_4 = 4.51$~K, which are
already quite close to the experimentally measured values. An even
better correspondence between the theoretical and the experimental
results for the three upper branches (shown by closed squares on
the $H=0$ axis of Fig.~\ref{fig4}) is achieved for a slightly
modified set of microscopic constants: $J=0.25$~K, $D_a=0.13$~K
($d_a=0.52$), and $E_d=0.55$~K ($d = 0.008$). The lowest gap in
this case is equal to $\Delta_1 = 1.13$~K, which is very close to
the estimate obtained by fitting the specific heat curve.

\section{Discussion}

\begin{figure}
\centerline{\includegraphics[width=\columnwidth]{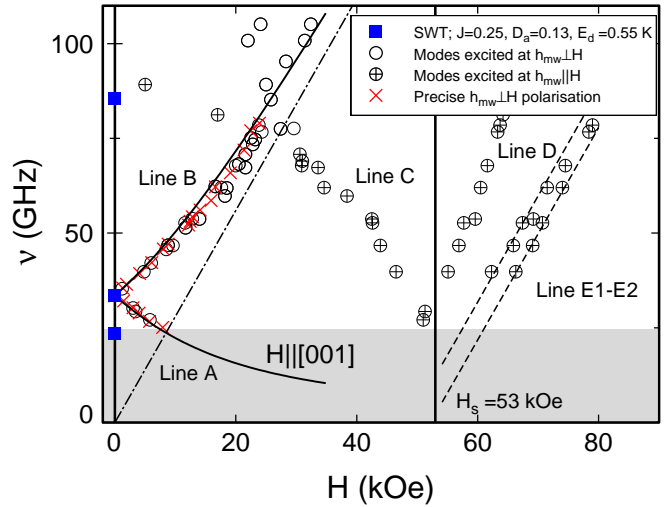}}
\caption{(Color online.) Frequency-field diagram of the resonance
spectrum observed on a powder sample of \GSO\ at $T=0.45$ K;
{\Large $\circ$} and $\oplus$ correspond to the spectra obtained
at ${\bf h}_{\rm mw}\perp {\bf H}$ and ${\bf h}_{\rm mw}\angle
{\bf H}$ respectively; $\blacksquare$ mark the gap values
calculated by the spin-wave theory (SWT); the solid lines
represent the solution of Eq.~(\ref{freq}) for a doublet mode with
$H\parallel [001]$, the dashed lines are linear fits to the two
components of spectral line E with $g=2$, and the dash-dotted line
corresponds to a $g=2$ paramagnet. The transition to a
spin-polarized phase at $H_s=53.0$\,kOe is marked by a vertical
line. The unreachable low-frequency range $\nu <25$~GHz is shaded
in grey.} \label{fig4}
\end{figure}

In the preceding section we have demonstrated a good agreement
between the measured ESR spectrum and the spin-wave calculations
performed in zero magnetic field. The two degenerate resonance
branches correspond to a peculiar type of spin motion: oscillation
of a spin plane with respect to two orthogonal in-plane axes. Such
modes are excited by the perpendicular component of the microwave
field ${\bf h}_{\rm mw}\perp {\bf H}$ as is indeed observed for
the lines~A and B.

Generalization to weak magnetic fields is straightforwardly done
with the ``hydrodynamic'' approach,\cite{marchenko}  which is
valid once the exchange interactions are significantly stronger
than magnetic anisotropies and field. Simple calculation analogous
to those performed in Ref.~\onlinecite{sosin1} yield the following
cubic equation for eigen-frequencies:
\begin{eqnarray}
& & (\nu^2 - \nu_1^2) (\nu^2 - \nu_2^2 )^2 \label{freq} \\ & &
\mbox{} \quad -\gamma^2\nu^2 \left
(\nu^2H^2-\nu_1^2H_{\parallel}^2 -\nu_2^2H_{\perp}^2 \right ) = 0,
\nonumber
\end{eqnarray}
where $\nu_1$ and  $\nu_2$  are resonance frequencies in zero
field, $H_{\parallel}$ and $H_{\perp}$ are magnetic field
components with respect to the tetragonal axis,
$\gamma=g\mu_B/2\pi\hbar$ is the electronic gyromagnetic ratio
($g=2.0$). In gadolinium stannate the magnetic anisotropies are comparable to
the nearest-neighbor exchange interaction. Moreover, the exchange
structure is soft, {\it i.e.}, infinitely degenerate, and the
anisotropies play a decisive role  in stabilizing the
observed magnetic structure. This restricts applicability of the
hydrodynamic theory, which is used here only to indicate a
plausible behavior.

The field evolution of the spectrum is summarized on the
frequency-field diagram presented in Fig.~\ref{fig4}. The two
degenerate modes appear to be split by the magnetic field into
decreasing (Line A) and increasing (Line B) branches. The points
in Fig.~\ref{fig4} mark the maximums of the ESR absorption which,
for a powder sample, correspond to one of the outermost field
orientations with respect to the crystal axes. The frequency-field
dependence of these maximums for the two degenerate modes is
satisfactorily fitted by formula (\ref{freq}) for $H\parallel
[001]$ if one takes $h\nu_2=\Delta_{2,3}$ (solid lines on
Fig.~\ref{fig4}), while the third calculated branch is field
independent and set to zero. When the orientation of the external
field is changed from $H\parallel [001] \rightarrow H\perp [001]$,
both branches shift to higher fields resulting in the overextended
right wings of the absorption lines observed in the experiment.
Hence, in spite of the ``softness'' of the exchange structure in a
Heisenberg pyrochlore magnet, the spin plane oscillations are not
strongly affected by quasi-local modes.

The third spectral mode~C is excited only by a parallel microwave
field component $h_{\rm mw}\parallel H$, which indicates that it
is not a uniform oscillation of the spin plane, but rather an
antiphase motion of spins of the cross (in-plane or out-of plane).
This mode should soften at the saturation point $H_s$ (as observed
in the experiment) and, on further increase of $H$, should develop
a gap, which probably corresponds to the increasing line~D of the
spectrum above $H_s$. The other two resonance branches (the
doublet line~E) observed at $H>H_s$ have linear field dependences
distinctive for quasi-local soft modes in the spin-polarized
phase.\cite{misha} They were also observed in \GTO\, but unlike
that case, one of them has almost zero energy at $H=H_s$, which
corresponds to the softening of one of the excitation branches at the
antiferromagnetic wave-vector ${\bf q}=0$ near the second-order
transition. (The ordering wave-vector in \GTO\ is different from
${\bf q} = 0$ and, therefore, all ESR modes remain finite at $H=H_s$.)
The spin-wave calculation of the high-field magnon spectrum and
its comparison with the observed results can provide a useful
information on the parameters of the the spin-Hamiltonian for the
two pyrochlore materials. Also, an interesting question remains
about possible phase transitions in Gd$_2$Sn$_2$O$_7$ at
intermediate fields $0<H<H_s$. Such transitions, if any, should be
quite sensitive to the field orientation, which hinders their
observation on the available powder samples.

\begin{figure}
\centerline{\includegraphics[width=\columnwidth]{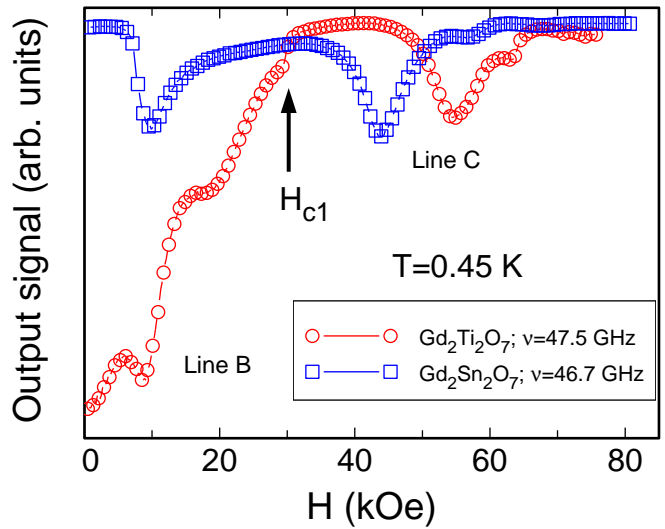}}
\caption{(Color online.) Absorption spectra recorded at nearly the
same frequencies in \GSO\ powder sample (squares) and in \GTO\
single crystal at $H\parallel [111]$ (circles); the arrow marks
the transition in \GTO\ at $H=H_{c1}\simeq 30$~kOe.} \label{fig5}
\end{figure}

The measured resonance modes and their
perfect agreement with the spin-wave calculations point at the
conventional magnetic ordering in \GSO\ below 1~K. Note, that an
explanation of the persistent spin dynamics observed by the local
probes \cite{bertin,dalmas,bonv1} remains an open issue: gapped magnons
with energies above 1~K cannot contribute to the muon spin
relaxation in the 50~mK range. Still, properties of the sister
material \GTO\ are strikingly different from the more conventional
stannate. This concerns not only the difference in the
low-temperature asymptotes for the specific heat,
\cite{gingras2,yaouanc} but also their ESR spectra as
illustrated in Fig.~\ref{fig5}. A typical absorption curve for
\GSO\ consists of several spectral lines with almost zero
background. In contrast, the recorded absorption curve for \GTO\
demonstrates a broad
intense non-resonant anomaly with a maximum in zero field. This
additional absorption develops simultaneously with the main spectral lines below
$T_N$, rapidly decreases with increasing magnetic
field and fully disappears stepwise at $H=H_{c1}\simeq 30$~kOe
marked by an arrow in Fig.~\ref{fig5}.
The above findings provide an evidence for additional
``magnetic degrees of freedom'' in the titanate, which exist down to
low temperatures and are suppressed by an external
magnetic field.

Additional low-energy excitations in \GTO\ may result from the
complexity of magnetic ordering in this material.
The  neutron diffraction experiments \cite{stewart} suggest a
multi-$k$ spin structure in the titanate, which may also lead to
multiple magnetic domains related to different combinations
of ${\bf q}=(1/2,1/2,1/2)$ and equivalent wave-vectors.
Excitations in the domain walls and their pinning by crystal defects
can produce additional low-temperature spin
dynamics. Such a residual dynamics (i) should be absent in the
${\bf q} =0$ ground-state of the stannate and (ii) exhibit a
significant sample dependence as was indeed found from the
comparison of $\mu$SR data obtained on single crystal
\cite{yaouanc} and powder samples.\cite{dunsiger} The
low-field  domain
structure can be further eliminated by a magnetic field,
which, for example, selects at $H>H_{c1}$ a unique ordering wave-vector.
An extra argument in favor of such a
scenario is that the ESR spectrum of the titanate is
significantly transformed at $H=H_{c1}$  becoming
similar to the spectrum of the stannate.\cite{sosin1} Namely, the
line~C of the spectrum, which is traced in the stannate in the
whole field range $0<H<H_s$, appears in the titanate only at $H>H_{c1}$.
High-field neutron diffraction measurements in \GTO\ together
with evolution of the low-$T$ asymptote in
the specific heat under magnetic field should provide a
valuable check for the above scenario.

In summary, the study of the magnetic resonance properties of the
pyrochlore gadolinium stannate reveals three gapped resonance
modes in the ordered phase, two of them being exactly degenerate
at zero external magnetic field.  The spin-wave theory,
which takes into account the nearest-neighbor exchange, the
single-ion anisotropy and the dipolar interactions, demonstrates
a very good agreement with the experiment using known values of the
microscopic magnetic parameters. The lowest gap value
predicted by theory lies beyond the experimental frequency range
and cannot be directly observed. Nevertheless, it roughly agrees
with an estimate made from the exponential decrease of the low
temperature specific heat.

\begin{acknowledgments}

The authors thank V. N. Glazkov and A. I. Smirnov for useful
discussions. Work at the Kapitza Institute is supported by RFBR
Grant 07-02-00725 and by the Program of the President of Russian
Federation. S.S.S. is grateful to INTAS for the financial support,
grant YSF 2004-83-3053 and to SPSMS/CEA-Grenoble for the
hospitality during the joint research program. Likewise, M.E.Z.
acknowledges warm reception at the Kapitza Institute.

\end{acknowledgments}

\end{document}